\documentclass{svjour3}                     
\smartqed  
\usepackage{graphicx}
\usepackage{mathptmx}      
\usepackage{latexsym}
 \journalname{Few-Body Systems}
\begin{document}

\title{Universal low energy features of two-body systems
}


\author{A. Calle Cord\'on         \and
        E. Ruiz Arriola.
}


\institute{A. Calle Cord\'on (Speaker) \at
           Departamento de F\'isica At\'omica, Molecular y Nuclear,
           Universidad de Granada, E-18071 Granada, Spain. \\ 
           \email{alvarocalle@ugr.es}
           \and
           E. Ruiz Arriola \at
           Departamento de F\'isica At\'omica, Molecular y Nuclear,
           Universidad de Granada, E-18071 Granada, Spain. \\
           \email{earriola@ugr.es}
}

\date{Presented at the 21st European Conference on Few-Body Problems in
Physics, \\Salamanca, Spain, 30 August - 3 September 2010}

\maketitle

\begin{abstract}
We apply renormalization ideas to study low-energy interactions in
two-body systems. As we will see this method highlights  a
model-independent description of a broad variety of systems ranging
from ultra-could atoms to NN and $\Lambda \Lambda$ interactions.

\end{abstract}

\vspace{1cm}

The renormalization with boundary conditions (BC) are best illustrated
by the effective range expansion where the S-wave phase shift is
given by ($p$ is CM momentum)
\begin{eqnarray}
p \cot \delta_0(p) = - \frac1{\alpha_0} + \frac12 r_0 p^2 + \dots 
\end{eqnarray}
where $r_0$ is the effective range and $\alpha_0$ the scattering
length. Using the superposition principle one obtains a bilinear
relation dubbed as {\it universal low energy
  theorem}~\cite{PavonValderrama:2005wv,RuizArriola:2007wm},
\begin{eqnarray}
r_{0} = A + \frac{B}{\alpha_{0}}+ \frac{C}{\alpha_{0}^2} \, ,
\label{eq:r0-a0}
\end{eqnarray}
where the coefficients $A$, $B$ and $C$ depend {\it only} on the long
distance potential. This way $V(r)$ and $\alpha_0$ are regarded as
{\it independent} variables.  We present below a variety of
situations.

\begin{figure}
\begin{center}
\includegraphics[height=5cm,width=4cm,angle=270]{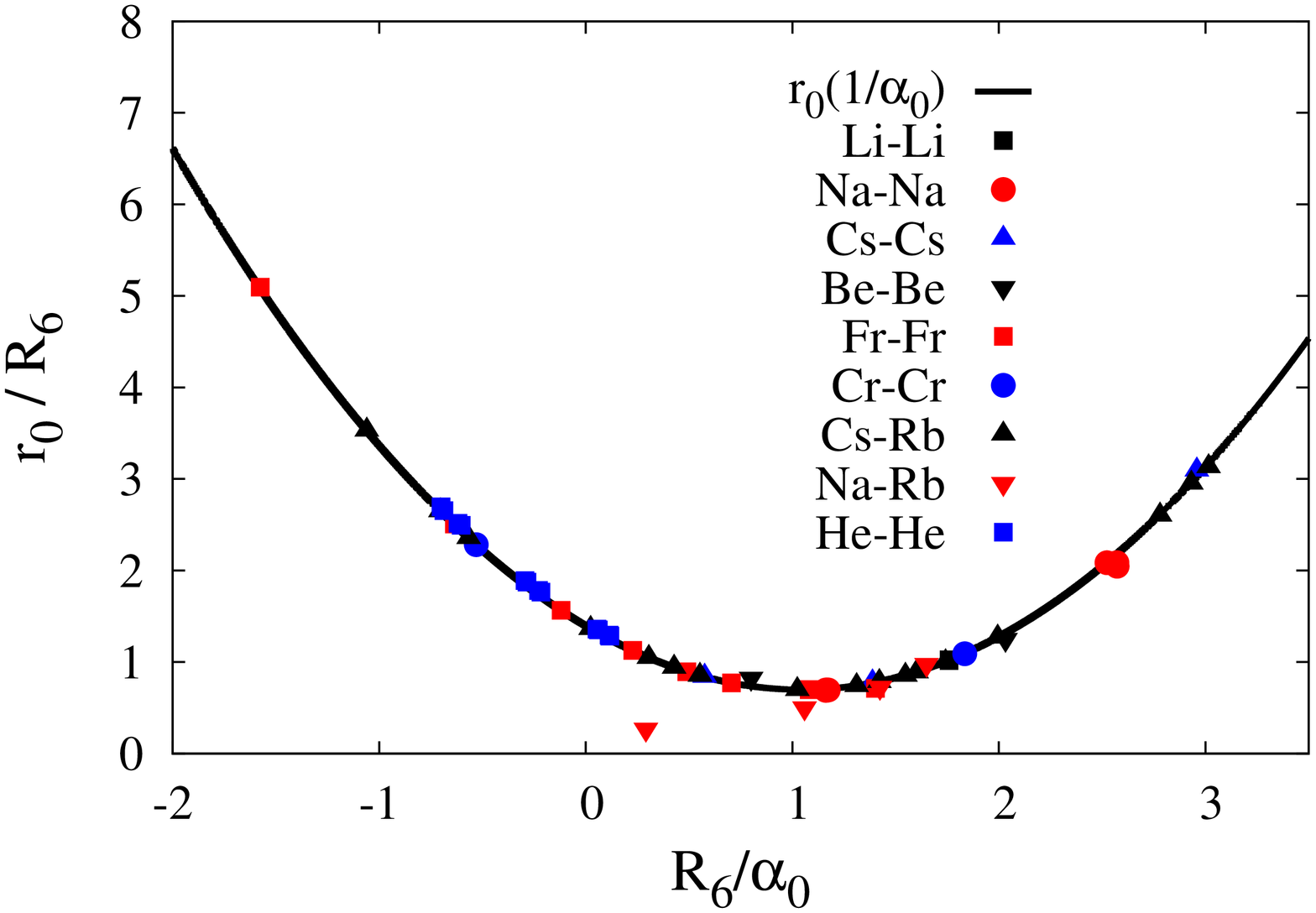}
\hspace{0.2cm}
\includegraphics[height=5cm,width=4cm,angle=270]{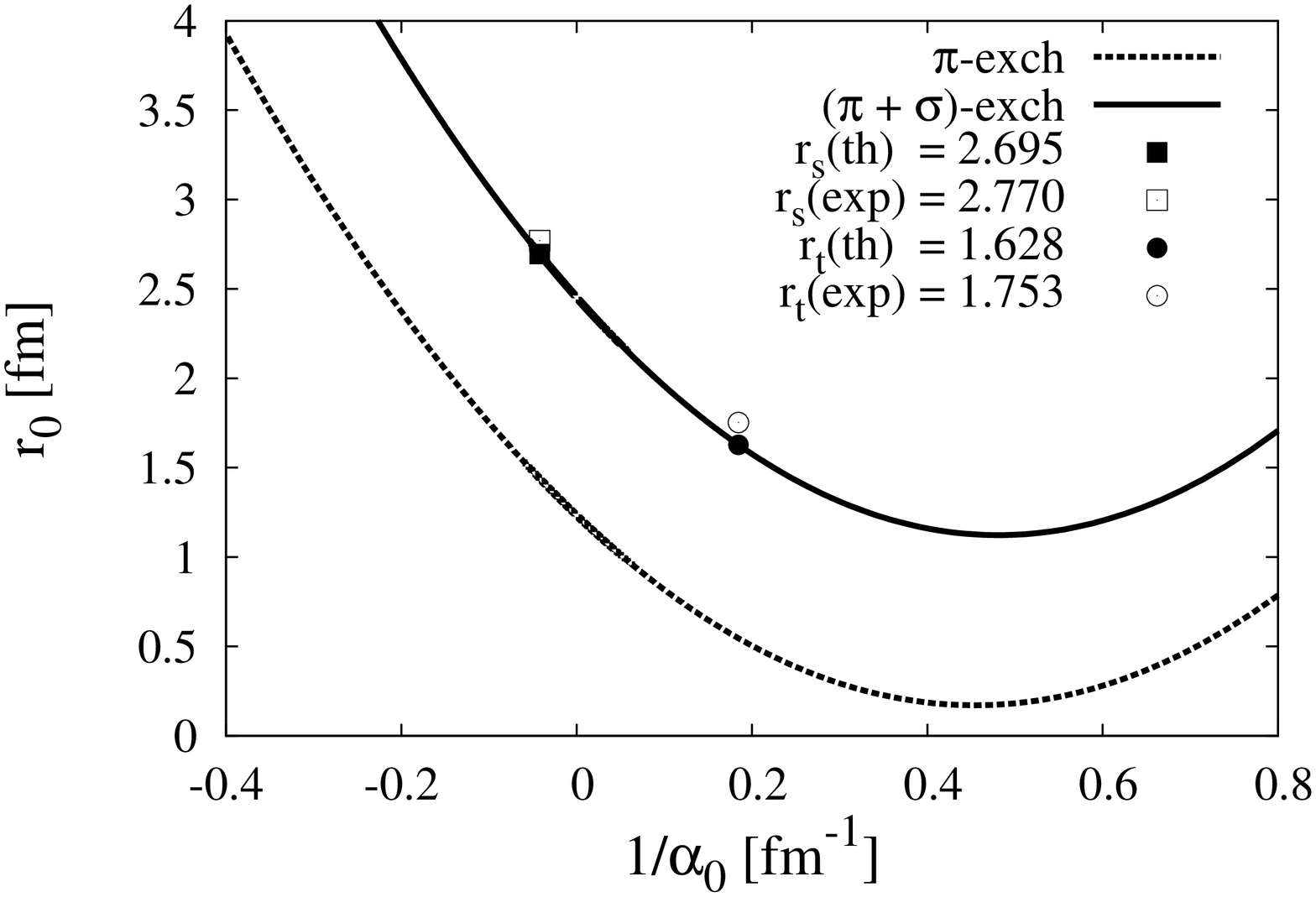}\\
\includegraphics[height=5.3cm,width=4cm,angle=270]{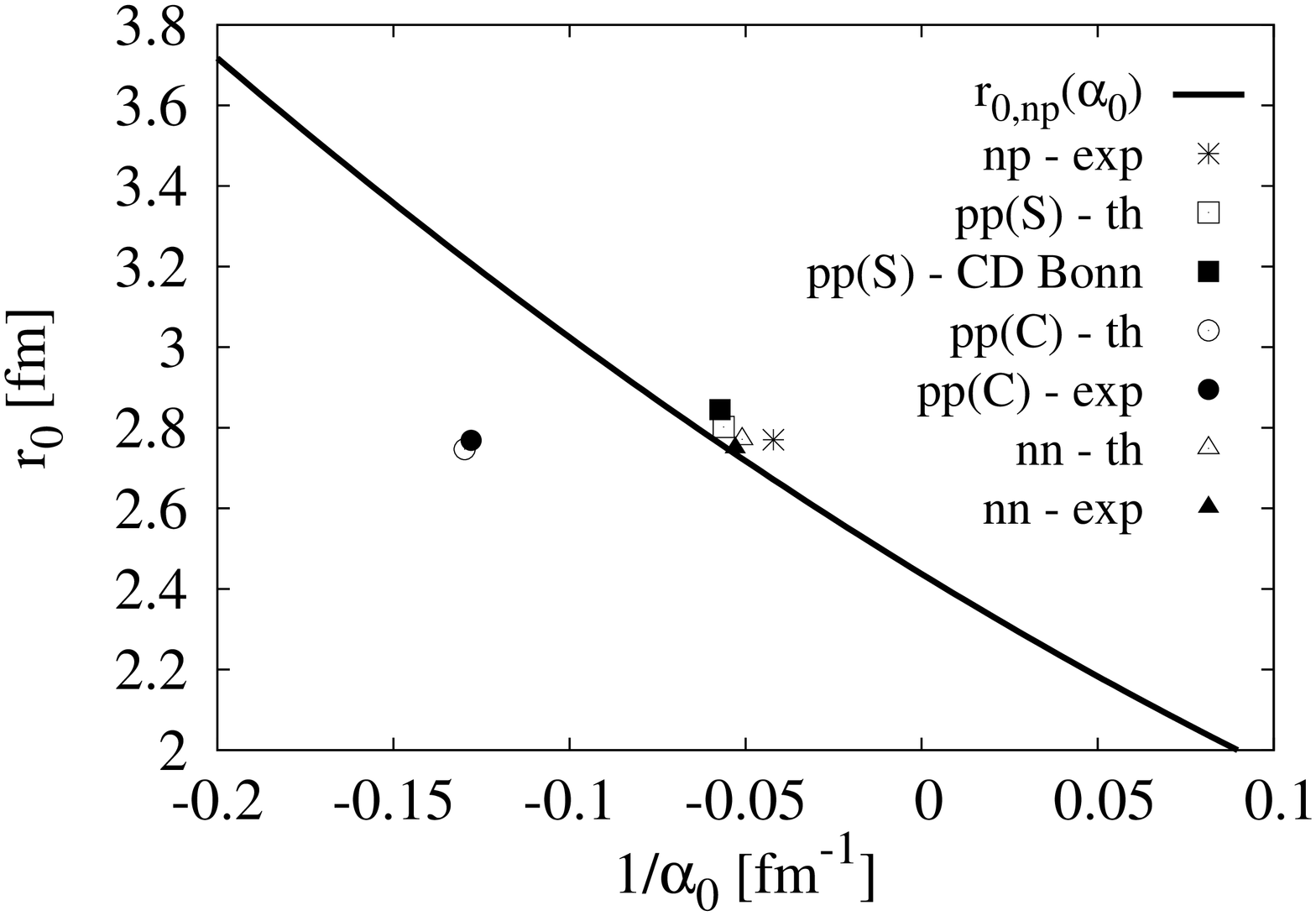}
\includegraphics[height=5.3cm,width=4.3cm,angle=270]{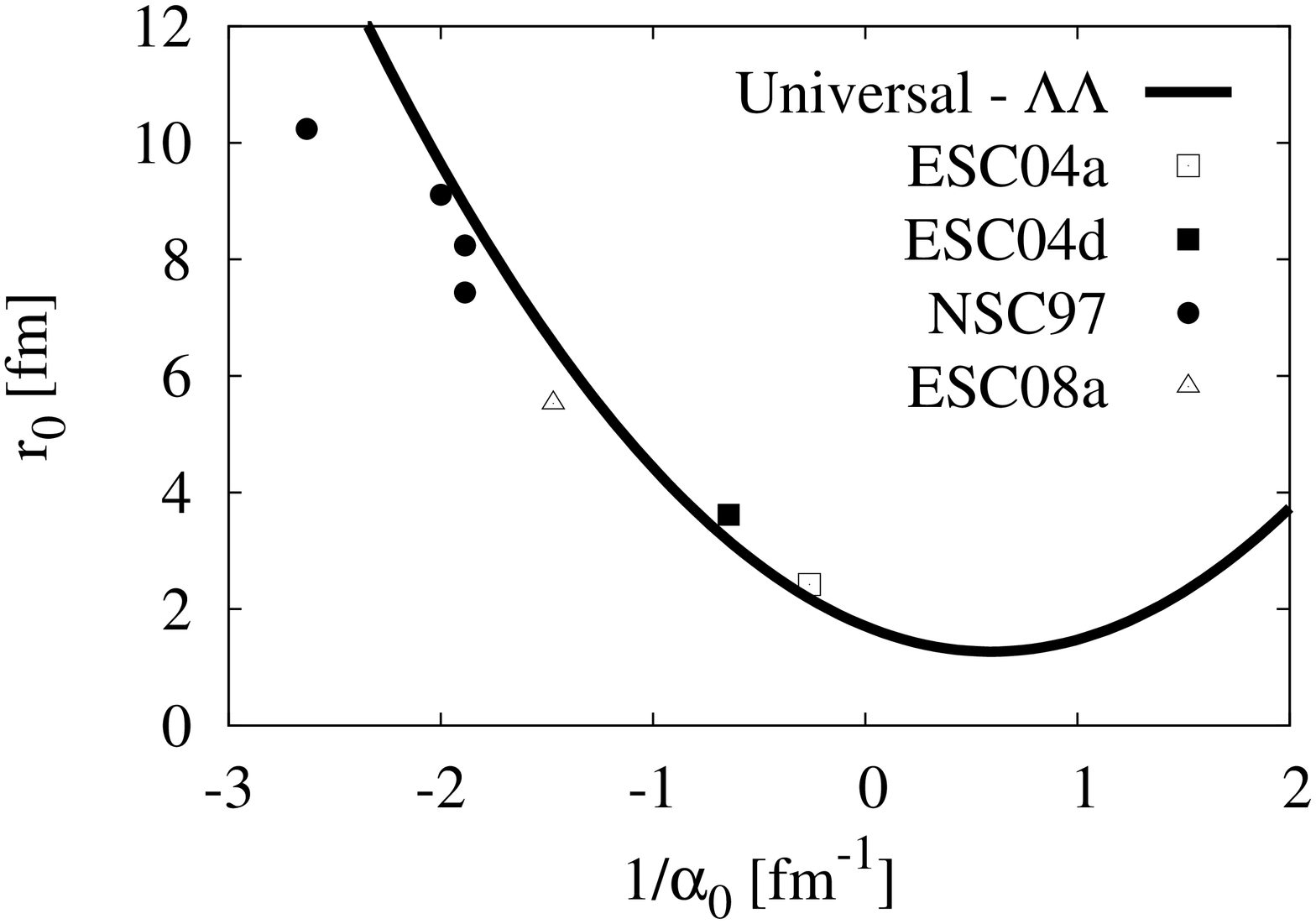}
\end{center}
\caption{Universal relations: (upper left panel) Atom-Atom vdW correlation in
  units of $R_6$, (upper right panel) NN $^1S_0$-$^3S_1$ Wigner
  correlation, (lower left panel) nn,np and pp(S) short-distance
  connection correlation and (lower right panel) $\Lambda\Lambda$ long
  distance correlation compared to potential models calculations (see
  e.g.  \cite{Yamamoto:2010zzi}) .}
\label{fig:r0-a0}      
\end{figure}

The case of ultra-could collisions between neutral atoms described by
the van der Waals (vdW) potential $V(r) = - \sum_{n=6}^{\infty}
\frac{C_n}{r^n}$ where $C_n$ are the vdW coefficients was analyzed in
Refs.~\cite{Arriola:2009wi,PhysRevA.81.044701}.  The terms $n=6,8,10$
are usually retained and the singularity appearing in the vdW
potential led to the use of phenomenological potentials, like the
Lennard-Jones, modeling short-distance physics.
The correlation, Eq.~(\ref{eq:r0-a0}), is plotted in
Fig.~\ref{fig:r0-a0} (upper left panel) for the term $n=6$ in units of
the vdW radius $R_6 = (2\mu C_6)^{1/4}$ where the coefficients
are $A/R_6 = 1.39473$, $B/R_6 = -1.33333$ and $C/R_6 = 0.637318$,
together with the experimental and potential model determinations of
$\alpha_0$ and $r_0$ known up to date and compiled in
Ref.~\cite{Arriola:2009wi,PhysRevA.81.044701}.
On the light of this clear universality one realizes that the leading $1/r^6$
contribution suffices to accurately describe low energy atom-atom scattering
with just two parameters in a wide energy range.
%


In our studies of the NN interaction we consider the OBE potential
with exchange of $\pi$, $\sigma$, $\rho$, $\omega$ keeping the
spin-flavor structure of the large-$N_c$ limit according with
Refs.~\cite{Kaplan:1996rk,Kaplan:1995yg} with the parameters fixed to
the condition that the $^1S_0$ $np$ phase shift be reproduced,
\begin{eqnarray}
V_{^1S_0}(r) =
 - \frac{g_{\pi NN}^2 m_\pi^2} {16 \pi M_N^2} \frac{e^{-m_\pi r}}{r}
 - \frac{g_{\sigma NN}^2}{4 \pi}\frac{e^{-m_\sigma r}}{r}
+ \frac{{g_{\omega NN}}^2}{4 \pi}\frac{e^{-m_\omega r}}{r} 
- \frac{{f_{\rho NN}}^2 m_\rho^2}{8 \pi M_N^2 } \frac{e^{-m_\omega r}}{r} \, .
\label{eq:OBE}
\end{eqnarray}
As was shown in Ref.~\cite{Cordon:2009pj} and more recently in
Ref.~\cite{Cordon:2010sq} the unnaturally large scattering length in this
channel triggers a fine-tuning problem in the potential parameters. However
using renormalization one disentangles the unknown short-distance physics from
the well established long-distance one by fixing from the start low-energy
parameters (LEPs).
In particular, one finds vector mesons to play a marginal role in the
description of NN scatering below pion production
threhold~\cite{Cordon:2009pj}. 
In addition 
one obtains
$m_{\sigma} = 501(25)\,{\rm MeV}$, $g_{\sigma NN} = 9(1)$,
 $g_{\omega NN} \sim g_{\omega NN}^{SU(3)} = 3 g_{\rho NN}$
even when charge symmetry breaking is implemented by
considering the neutron-proton and charged-neutral pion mass
differences~\cite{Cordon:2010sq}.
%
Using the more conventional regular boundary condition $u(0) = 0$ two
very good but mutually incompatibles fits are obtained, two scenarios
which correspond with the absence or presence of spurious bound
states~\cite{Cordon:2009pj} and in any case with or without large
$SU(3)$ violation of $g_{\omega NN} $ respectively. 

Using the previous potential in Ref.~\cite{CalleCordon:2008cz} we have
re-interpreted the old nuclear Wigner (spin-isospin) SU(4) symmetry as
a {\it long distance symmetry}.
This symmetry implies equal singlet and triplet potentials $V_{1S0} =
V_{3S1}$, a fact that agrees with the large-$N_c$ expectations.
The renormalization approach allows a breaking of the symmetry at
short distances having $\alpha_{0,1S0} \neq \alpha_{0,3S1}$ and $r_{0,1S0}
\neq r_{0,3S1}$ even with the same long-distance potential. This can be
clearly seen in Fig.~\ref{fig:r0-a0} (upper right panel) where we plot
Eq.~(\ref{eq:r0-a0}) in the case of OPE and OPE+$\sigma$. The only
difference between $r_{0,1S0}$ and $r_{0,3S1}$ resides in the
numerical values of the scattering lengths $\alpha_{0,1S0}$ and
$\alpha_{0,3S1}$ but the coefficients $A$,$B$ and $C$ in
Eq.~(\ref{eq:r0-a0}) are the same in both channels. The experimental
values fall almost on top of the curve.

The posibility of connecting different NN isospin channels has been
analyzed in Ref.~\cite{Cordon:2010sq} using a {\it short distance
  connection}. One is able to obtain a bilinear relation between the
scattering lengths of different problems $ \alpha_{0,2} =
(a\ \alpha_{0,1} + b)/ (c\ \alpha_{0,1} + d) $ where the coefficients
$a$,$b$,$c$ and $d$ depend {\it only} on the long distance
potential. Fixing the scattering length in one channel,
e.g. $\alpha_{0,np} = -23.74\,{\rm fm}$, with the suitable
incorporation of the Coulomb interaction $V_C(r) = \alpha/r$, with
$\alpha \simeq 1/137$ for the pp(C) case, we obtain the rest of LEPs
displayed in Table~\ref{tab:ABC-CSB}.
\begin{table}
\centering
\begin{tabular}{c|ccc|cc|cc}
\hline
BB    & A [${\rm fm}$] & B [${\rm fm^2}$] &  C [${\rm fm^3}$]
      &  $\alpha_0 [ {\rm fm} ] $ & $r_0 [ {\rm fm} ] $ 
      &  $\alpha_0^{\rm exp} [ {\rm fm} ] $ & $r_0^{\rm exp} [ {\rm fm} ] $ 
\\\hline
np    & 2.437 & -5.345 & 5.277 & input   & 2.672  & -23.74(2) & 2.77(5)
\\
pp(S) & 2.465 & -5.661 & 5.996 & -17.806 & 2.802  & -17.46    & 2.845\\
pp(C) & 1.982 & -4.985 & 7.009 &  -7.706 & 2.747  &  -7.8149(29) & 2.769(14) \\
nn    & 2.467 & -5.666 & 6.003 & -19.626 & 2.771  & -18.9(4)  & 2.75(11) \\
\hline
\end{tabular}
\caption{Coefficients $A$,$B$ and $C$ of Eq.~(\ref{eq:r0-a0}) using
  the OBE potential Eq.~(\ref{eq:OBE}) implementing CSB and values for
  the LEPs using the short distance connection. We also list
  experimental or recommended values.}
\label{tab:ABC-CSB}
\end{table}
In Fig.~\ref{fig:r0-a0} (lower left panel) we have plotted
Eq.~(\ref{eq:r0-a0}) for the np system and the predicted LEPs for the
other channels. This figure indicates that the strongly interacting
component is close to a unique curve whereas adding the
Coulomb interaction distortes the long distance potential and hence
the pp(C) LEPs move away from the curve. No explicit $\rho-\omega$
mixing seems necessary as it is a short range effect which is
reparameterized in terms of $\alpha_0$.

Finally, we can assume systems with strangeness such as
$\Lambda\Lambda$ for example (see e.g.  \cite{Yamamoto:2010zzi} and
references therein). If $\Lambda$-hyperons interact exchanging
$1\sigma$- and $1\omega$-mesons and we use SU(3) relations for the
couplings $g_{\sigma \Lambda\Lambda} = g_{\omega NN}$ and $g_{\omega
  \Lambda\Lambda} = 2/3 g_{\omega NN}$ we obtain $A= 1.705 {\rm fm} $,
$B= -1.475 {\rm fm}^2$ and $ C=1.245 {\rm fm}^3$. The universal curve
for such a simple OBE potential is displayed in Fig.~\ref{fig:r0-a0}
(lower right panel) and is not far from more sophisticated potential
model calculations~\cite{Yamamoto:2010zzi}.

\begin{acknowledgements}
We thank M. Pav\'on Valderrama for his collaboration on
Ref.~\cite{Cordon:2010sq}.  This work is supported by the Spanish DGI
and FEDER funds with grant FIS2008-01143/FIS, Junta de Andalucía grant
FQM225-05. 
\end{acknowledgements}


\end{document}